# A boron-coated CCD camera for direct detection of Ultracold Neutrons (UCN)

K. Kuk[1], C. Cude-Woods[2,3], C. R. Chavez[1], J. H. Choi[3], J. Estrada[1], M. Hoffbauer[2], M. Makela[2], P. Merkel[1], C. L. Morris[2], E. Ramberg[1], Z. Wang[2] , T. Bailey[3], M. Blatnik[4], E. R. Adamek[5], L. J. Broussard[6], M. A.-P. Brown[7], N. B. Callahan[8], S. M. Clayton[2], S. A. Currie[2], X. Ding[9], D. Dinger[10], B. Filippone[4], E. M. Fries[4], P. Geltenbort[11], E. George[10], F. Gonzalez[8], K. P. Hickerson[4], K. Hoffman[10], A. T. Holley[10], T. M. Ito[2], S. W. T. MacDonald[2], C.-Y. Liu[8], C. O'Shaughnessy[2], R. W. Pattie, Jr.[12], D. G. Phillips II[3], B. Plaster[7], D. J. Salvat[8], A. Saunders[2], S. Seestrom[13], E. I. Sharapov[14], S. K. L. Sjue[2], V. Su[4], X. Sun[4], Z. Tang[2], P. L. Walstrom[2], W. Wei[4] , J. W. Wexler[3], T. L. Womack[2], A. R. Young[3],  B. A. Zeck[3] and UCN collaboration

Contact: Zhehui Wang, zwang@lanl.gov; Juan Estrada, estrada@fnal.gov

[1]Fermi National Accelerator Laboratory, Batavia, IL 60510, USA
[2]Los Alamos National Laboratory, Los Alamos, NM 87545, USA
[3] North Carolina State University, Raleigh, North Carolina 27695, USA
[4] California Institute of Technology, Pasadena, California 91125, USA
[5] National Institute of Standards and Technology, USA
[6] Oak Ridge National Laboratory, USA
[7] University of Kentucky, Lexington, Kentucky 40506, USA
[8] Indiana University, Bloomington, Indiana 47408, USA
[9] Virginia Polytechnic Institute and State University, USA
[10] Tennessee Technological University, USA
[11] Institut Laue-Langevin, France
[12] East Tennessee State University, USA
[13]Sandia National Laboratories, Albuquerque, NM 87123, USA.
[14]Joint Institute for Nuclear Research, Russia.
(Feb. 24, 2019)

**Abstract**: A new boron-coated CCD camera is described for direct detection of ultracold neutrons (UCN) through the capture reactions $^{10}B$ (n,α0γ)$^7Li$ (6%) and $^{10}B$(n,α1γ)$^7Li$ (94%). The experiments, which extend earlier works using a boron-coated ZnS:Ag scintillator, are based on direct detections of the neutron-capture byproducts in silicon. The high position resolution, energy resolution and particle ID performance of a scientific CCD allows for observation and identification of all the byproducts α, $^7Li$ and γ (electron recoils). A signal-to-noise improvement on the order of $10^4$ over the indirect method has been achieved. Sub-pixel position resolution of a few microns is demonstrated. The technology can also be used to build UCN detectors with an area on the order of 1 $m^2$. The combination of micrometer scale spatial resolution, few electrons ionization thresholds and large area paves the way to new research avenues including quantum physics of UCN and high-resolution neutron imaging and spectroscopy.

## Introduction

Ultracold neutrons (UCN) are free neutrons with a kinetic energy less than about 300 neV. UCN have been used to examine fundamental interactions in nature through precise determination of neutron lifetime [1] [2], determination of coefficients of neutron beta-decay [3], search for neutron electric dipole moment [4], etc. Recently, there is growing interest in using position-sensitive measurement of UCN to study fundamental quantum states. For example, quantum states of UCN in the Earth's gravity field have been reported in 2002 [5]. Precise measurement of quantum states of UCN can be used in dark energy and dark matter [6] [7]. Gravitationally bound UCN might also be used to test Newton's Inverse Square Law of gravity at short distances from 0.1-100 µm [8][9][10]. Most of the gravity quantum states measurements require a spatial resolution less than 10 µm, and 1 µm or less is highly desired.

## Charge Coupled-Device (CCD) and thin $^{10}$B film for neutron detection

The CCDs used in this work are engineering grande sensors similar to those in the DECam wide field imager [11]. The CCDs have been built by the Lawrence Berkeley National Laboratory (LBNL) [12], and extensively characterized at Fermilab for the DECam project [13]. The DECam CCDs are 250 µm thick, fully depleted, back-illuminated devices fabricated on high-resistivity silicon. The CCDs used for this study have 8 million pixels with dimensions 15 µm x 15 µm. **Figure 1** shows the 3-phase, p-channel CCD design. The 10 kΩ-cm resistivity, allows for a fully depleted operation at bias voltages of 25 V. The field extends essentially all the way to the backside contact, depleting the entire volume of the CCD substrate.

When an ionizing particle penetrates the detector, it creates electron–hole pairs. Under the influence of the electric field, the holes produced near the back surface will travel the full thickness of the device to reach the potential well near the gates. Particle identification for this detector has been discussed in previous work [14] with special attention paid to identifying heavily ionizing alpha particles, as discussed in Ref. [15]. These alpha particles, such as those expected from the neutron capture reactions $^{10}$B $(n,\alpha0\gamma)^{7}$Li and $^{10}$B$(n,\alpha1\gamma)^{7}$Li are easily identified in CCD images thanks to the plasma effect.

These CCDs have been used before for thermal neutron detection using a 1 µm thick boron film in a ceramic substrate that was positioned close to the CCD, as discussed in Ref.[16]. For this work, the back of the CCD itself was directly coated onto the CCDs with $^{10}$B to less than 100 nm thick, using the same electron-beam evaporation method described previously [17]. This boron-coated CCD (bCCD) sensor becomes then ideal for detecting UCN with high position resolution. This is an extension of recent work using boron-coated ZnS:Ag scintillator for indirect detection of UCNs [18] [19].

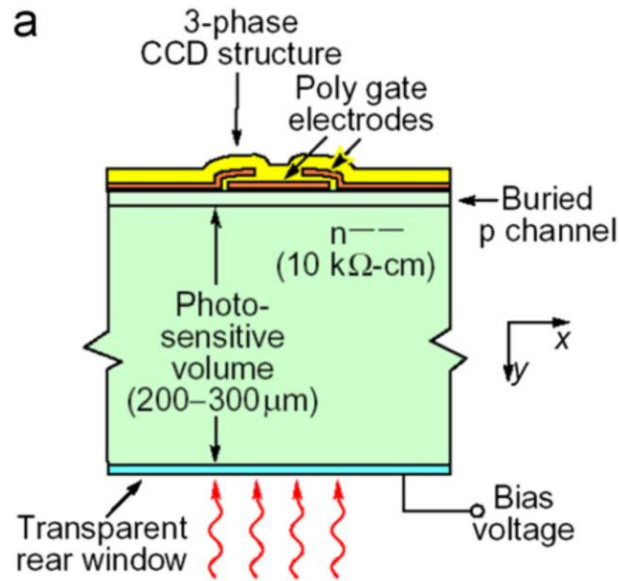

*Figure 1. Pixel cross section of a 250 um thick CCD developed at Lawrence Berkeley Laboratory.*

## Experimental setup at the LANSCE UCN facility

The performance of the bCCD as a UCN detector was tested at the LANSCE UCN facility. This UCN source uses solid deuterium to cool spallation neutrons by 15 orders of magnitude [20] [21]. The resulting UCNs move at speeds of only a few meters per second, and are completely confined by magnetic fields and/or material.

The bCCD was installed inside a vacuum vessel, and cooled to 140K to suppress dark current. This vessel was then connected to the UCN source as shown in **Figure 2**. The sensor was positioned with the borated surface facing up, and directly exposed to the UCN flux.

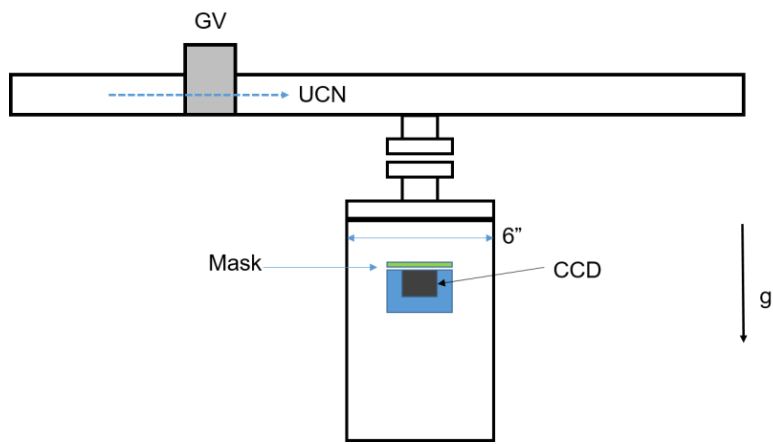

Figure 2. *The schematic of the experimental setup. UCN flux is controlled by a gate valve (GV). The CCD is facing up. Experiments are conducted both with and without a mask.*

**Position resolution.**

The spatial resolution of the UCN-CCD was demonstrated using two different targets. The first target was a 3D printed on thermoplastic material. The target was designed to sit a ~ 3mm away from the borated surface of the bCCD, and covered completely the imaging sensor. The detector was then exposed to UCN during which the hits were recorded in every CCD channel. The pattern of the mask was reproduced quite well by the hits in the CCD, as shown in **Figure 3**. This initial test qualitatively demonstrates the imaging capability of the sensor, and is also a good demonstration of the particle identification discussed in Ref. [18]. It is clear that the bCCD sees heavily ionizing particles (alpha and Li ions) and also compton scattering from gammas produced in the $^{10}B$ (n,α0γ)$^7Li$ (6%) and $^{10}B$(n,α1γ)$^7Li$ (94%) reactions.

A more quantitative measurement of the position resolution was performed with a second mask. This mask was built with 3 layers: (a) copper 2 mm , (b) double-sided sticky tape 80 um, (c) Polymethylpentene film (trademark TPX)  which came in two different thicknesses of 0.5 mm and 80 um. These layers were arranged as shown in **Figure 4**. A circular hole was punched through the complete assembly. In this case the mask was brought into direct contact with the borated surface of the bCCD. Two similar masks were built and installed next to each other to cover the full imaging detector as shown in **Figure 4**. The resulting hit map *[about 5 minutes of beam]* is shown in **Figure 5.**  Here, only the hits consistent with the circular charge clusters produced by heavily ionizing particles were selected, removing all the tracks from compton electrons.

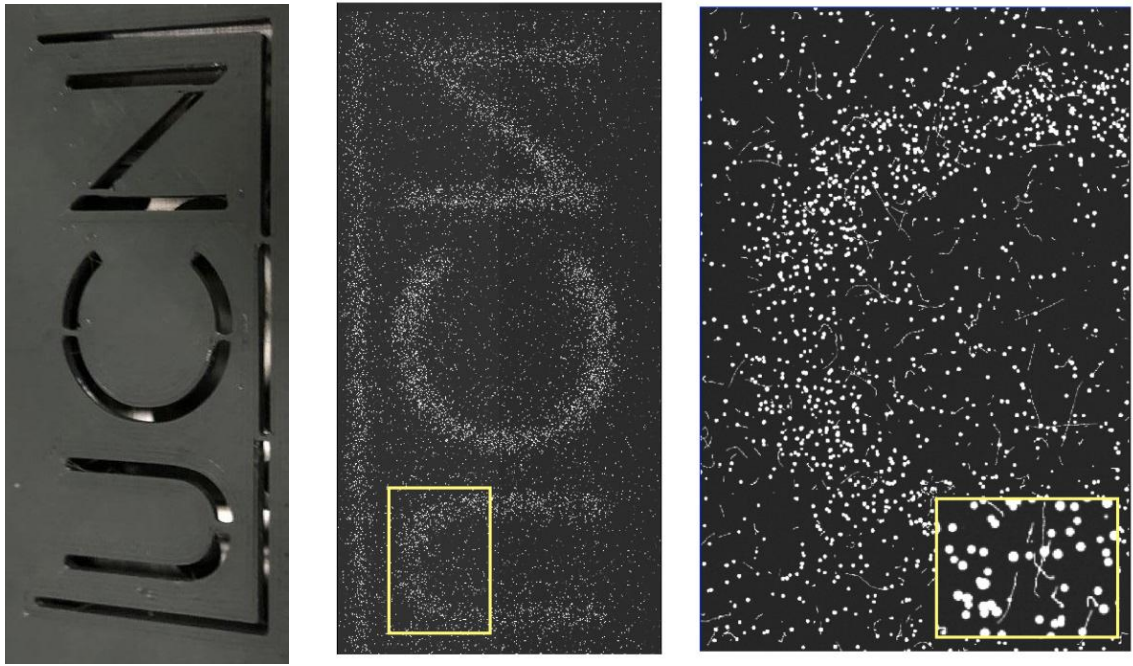

Figure 3.  *(Left) A 3D printed mask using thermoplastic material. This mask covered the full size of the detector (6cm x 3cm). The mask was positioned about 3mm from the borated surface of the bCCD. (center) raw bCCD image showing the transmission flux of UCN through the mask. (right) Zoom into the yellow box at the center panel, showing the hits from the different particles. Inset shows a zoom into an even smaller region of the image.*

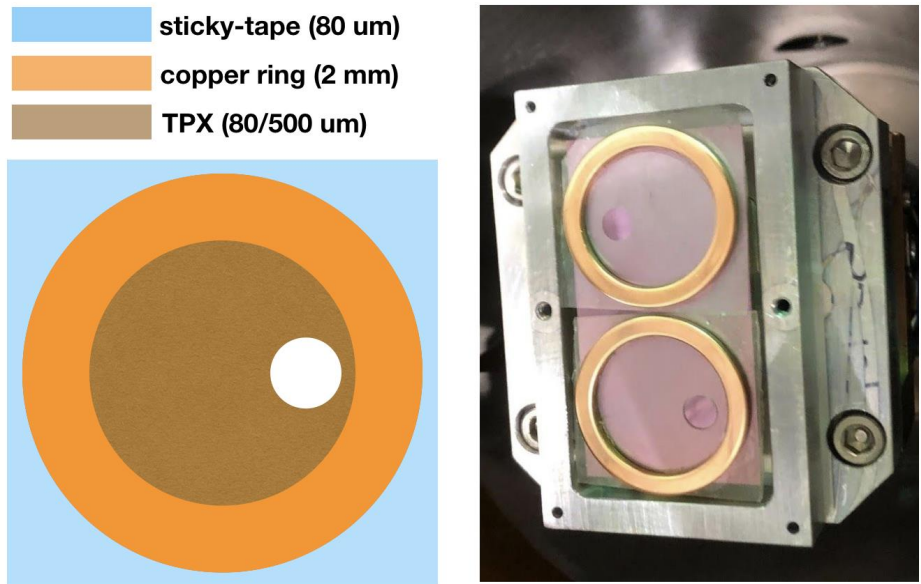

*Figure 4. (left) Schematic of the three layer mask using in from of the bCCD. Two versions of this mask were used, with different thickness of the kapton tape (80 $\mu m$ and 500 $\mu m$ ). (right) Photograph of the two masks directly mounted in contact with the borated surface of the bCCD.*

The smaller circle in the lower right in **Figure 5** is used for the estimation of the position resolution. The centroid for the circle is calculated, and the distribution of hits per unit area is shown for that circle in **Figure 6**. This distribution is fitted to the convolution of a gaussian resolution with a step function, and the results of the fit are shown in **Figure 6**, indicating a gaussian resolution of 13 pixels (195 um). It should be noted that the mask was not cut with micron level precision for this initial experiment, and part of this resolution can be attributed to the mask fabrication. To demonstrate the issue of the mask imperfection, a new fit was performed using only the upper right quadrant of the circle giving a resolution of 4 pixels (60 um).

The result shown in **Figure 6** demonstrates good position resolution, but it is likely limited by imperfections on the mask, and not by the intrinsic position resolution of the detector. The hits produced by heavily ionizing particles are extended over many pixels, and their centroid can be determined with sub-pixel precision. To demonstrate this we fit the charge distribution produced by a single highly ionizing hit reconstructed on the bCCD image. As shown in **Figure 7**, the centroid of each hit can be determined with an uncertainty significantly smaller than a pixel. The fit shown in **Figure 7** has an uncertainty of 0.1 pixel (1.5 um). The intrinsic resolution of the detector will be determined by the range of the alpha particles inside the silicon, as discussed in Ref. [16] This range is expected to be no more than 6.4 um as shown in Table.1. The energy loss in the $^{10}$B layer and the dead layer on the surface of the silicon active region will reduce the actual stopping ranges of the ions in silicon.

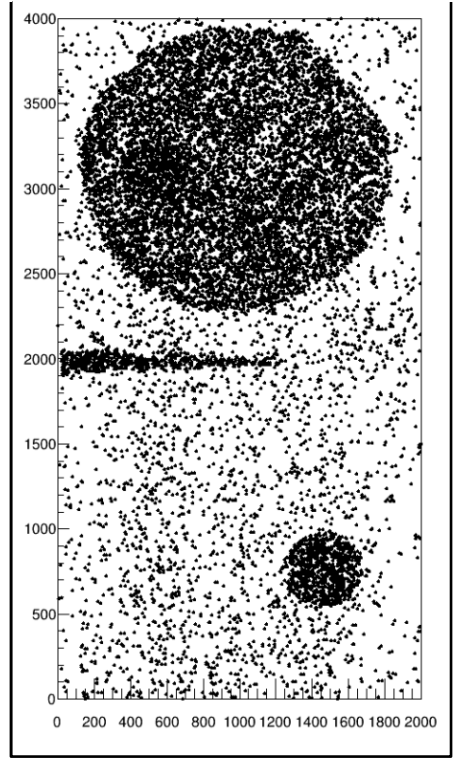

Figure 5. *Hit map for the observed heavily ionizing particles for the setup discussed in* Figure 4. *The circles produced by the different layers of the mask are clearly reconstructed. The hit map also shows the triangular gap between the two masks.*

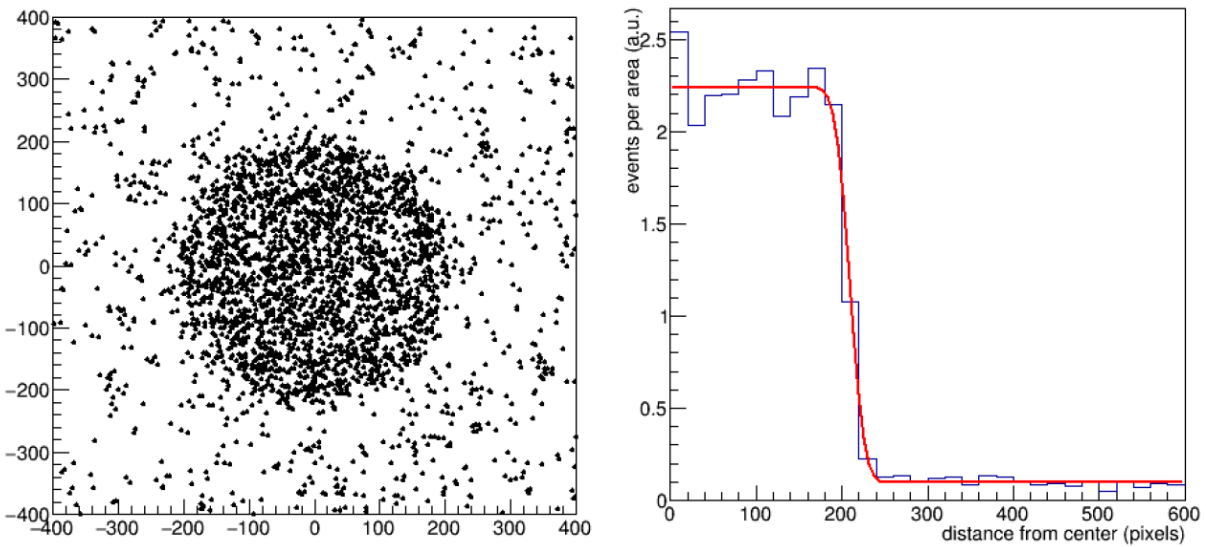

Figure 6. *(left) Hit map for heavily ionizing particles reconstructed on the bCCD image for the setup in* Figure 4. *This is a subsection of the hit map shown in* Figure 5, *with the horizontal and vertical scale centered at the circular feature in the bottom right of* Figure 5. *(right) The number of hits per unit area is shown as a function of distance from the center of the circle. The red curve shows the fit to a step function with a gaussian position resolution as discussed in the text.*

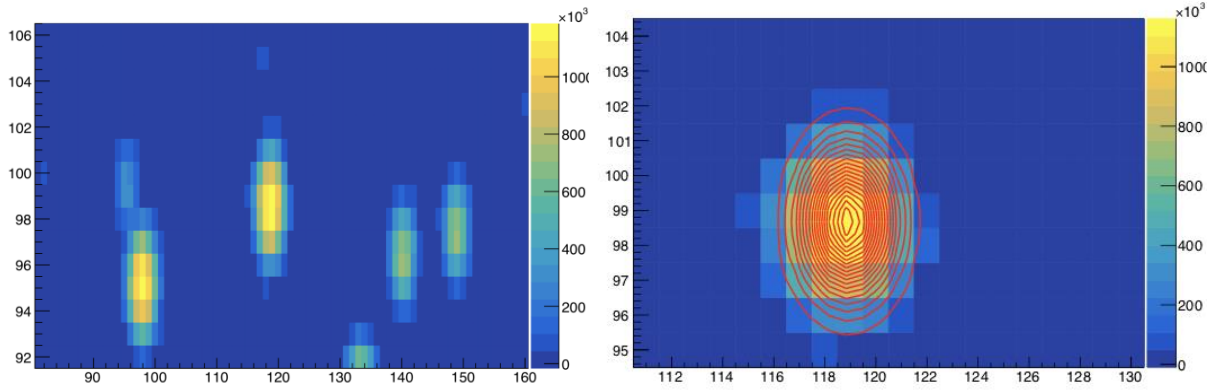

*Figure 7. (left) Small region of reconstructed bCCD image with 5 heavily ionizing hits, both axis are in pixels. Color scale is in digital units, corresponding to 40 couts/electron. (right) Zoom in the region around column 120. The red contours correspond to a 2D gaussian fit to the hit, demonstrating the determination of the centroid with sub pixel resolution.*

*Table 1. Maximum ion ranges ($R^i$) and their detection probabilities ($w^i$) of the charged products from the capture reactions $^{10}B$ ($n,\alpha 0\gamma$)$^7Li$ (6%) and $^{10}B(n,\alpha 1\gamma)^7Li$ (94%) in $^{10}B$ layer and Si sensor. A possible dead layer may exist in-between the 10B layer and active region of Si sensor, which could reduce the actual range in Si sensor.*

| Ion (probability, $w^i$) | Energy ($E_0^i$, MeV) | Range in $^{10}B$ ($R^i$, µm) | Range in Si ($R^i$, µm) |
|---|---|---|---|
| $\alpha$ (47%) | 1.47 | 3.5 | 5.1 |
| $\alpha$ (3%) | 1.78 | 4.4 | 6.4 |
| $^7Li$ (47%) | 0.84 | 1.8 | 2.5 |
| $^7Li$ (3%) | 1.02 | 2.1 | 2.8 |

**Energy reconstruction.**

The CCD detector was calibrated using an $^{55}$Fe radiation source producing mainly 5.9 keV X-rays, as discussed in Ref [13]. This calibration was then used to measure the energy of the events consistent with a heavily ionizing particle resulting from the neutron capture. The energy spectrum for the heavily ionization events observed during the UCN exposure is shown in **Figure 8.** The expected peaks from the reaction $^{10}B(n,\alpha 1\gamma)^7Li$ are 1.47 MeV and 0.84 MeV for $\alpha$ and $^7Li$ respectively. The position of the energy peak for the $\alpha$ is properly reconstructed, while the $^7Li$ is shifted to about 0.7 MeV. A more detailed study of the reconstruction will have to be performed to understand this shift. However, it is clear that it is possible to separate both type of events, allowing possible achievable position resolution for single neutron capture events to be below the upper limit of 6.4 um.

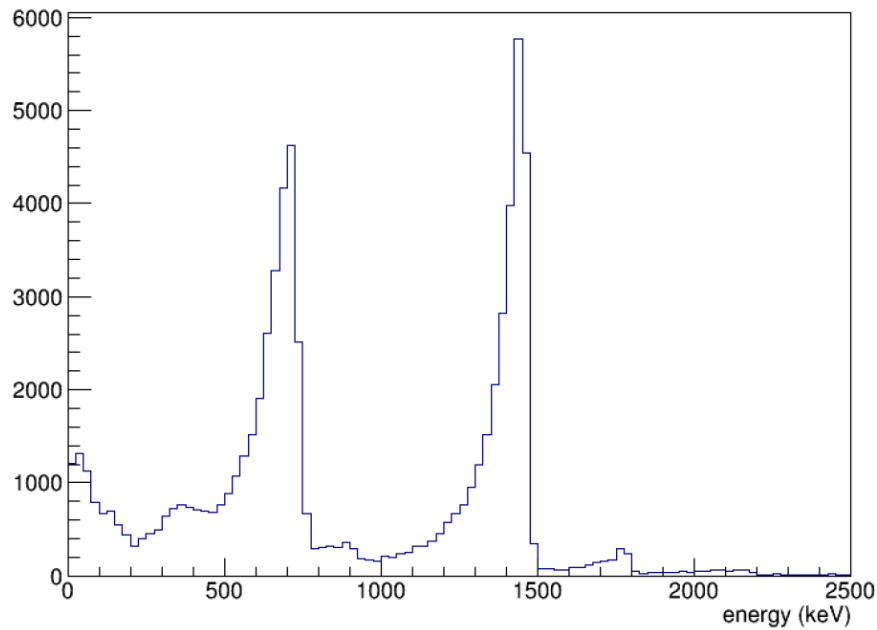

Figure 8. *UCN capture spectrum as measured by the calibrated bCCD (expected: alpha=1.47 MeV, Li=0.84 MeV).*

**Conclusion**

The detection of UCN was demonstrated at the LANSCE UCN facility, using a back illuminated CCD with a layer of approximately 100 nm of $^{10}$B (bCCD) directly deposited on the active surface. The energy spectrum of the hits in this detector shows that it is possible to separate the products of the neutron capture (α and $^7$Li), and the particle ID for the detector allows to identify the Compton events produced from gamma radiation. A position resolution of 60 um was demonstrated using a mask installed directly on top of the bCCD. This is still far from the ultimate position resolution estimated for this technique. The interpretation of the TPX edge is complicated by the penumbra produced because of UCN absorption in the edge of the plastic above the surface, which is thick compared to the expected resolution of the detector. This effect can be mitigated in future tests by using a reflective rather than an absorptive material in contact with the CCD.

The bCCD detector demonstrated in this work provides a new technique for the measurement of ultra cold neutrons with high position resolution. Several ongoing and planned experiments could benefit from the performance shown here such as real-time detector in studying UCN quantum states in the Earth's gravitational field. Other possible applications include UCN microscopy and reflectometry for material science as discussed in Ref[18]. Opportunities to examine the quantum properties of UCNs, building on demonstrations of quantum physics under gravity, have also been recognized by the community report in Ref [22]. Current UCN experiments with position measurements have been performed and planned as a probe of dark energy models [23][24][25]. The greatly enhanced UCN position and energy resolution from the device discussed here could enhance the capability of these experiments.

**References**

[1] A. P. Serebrov, E .A. Kolomensky, A. K. Fomin, I. A. Krasnoschekova, A. V. Vassiljev, D. M. Prudnikov, I. V. Shoka, A. V. Chechkin, M. E. Chaikovskiy, V. E. Varlamov, S. N. Ivanov, A. N. Pirozhkov, P. Geltenbort, O. Zimmer, T. Jenke, M. Van der Grinten, M. Tucker, "Neutron lifetime measurements with the big gravitational trap for ultracold neutrons", Phys. Rev. C 97, 055503 (2018)

[2] R. W. Pattie et al, 'Measurement of the neutron lifetime using a magneto-gravitational trap and in situ detection.' *Science* 360, 627 (2018).

[3] M. P. A. Brown et al, 'New result for the neutron -asymmetry parameter from UCNA.' Phys. Rev. C 97, 035505 (2018)

[4] B. W. Filippone, 'Worldwide search for the neutron EDM.' arXiv:1810.03718.

[5] Nesvizhevsky, V. V. et al. "Quantum states of neutrons in the Earth's gravitational field". Nature 415, 297 - 299 (2002).

[6] T. Jenke, G. Cronenberg, M. Thalhammer, T. Rechberger, P. Geltenbort, H. Abele, " Gravity experiments with ultracold neutrons and the qBounce experiment", arXiv:1510.03078 (2015)

[7] Guillaume Pignol et al, "Probing Dark Energy Models with Neutrons. Physics", Int. J. Mod. Phys. A, 30, 1530048 (2015)

[8] J. S. Nico, W. M. Snow, "Experiments in Fundamental Neutron Physics", Ann.Rev.Nucl.Part.Sci.55:27-69,2005

[9] Abele, 2008, "The neutron. Its properties and basic interactions", Progress in Particle and Nuclear Physics, Volume 60, Issue 1 (2008)

[10] D.Dubbers and M. G. Schmidt,"The neutron and its role in cosmology and particle physics", Rev. Mod. Phys. 83, 1111 (2011).

[11] B. Flaugher et al, "The Dark Energy Camera", The Astronomical Journal, Volume 150, Issue 5, article id. 150, 43 pp. (2015).

[12] Holland, S.E. et al.: Fully depleted, back-illuminated charge-coupled devices fabricated on high-resistivity silicon, IEEE Trans. Electron Devices 50(1), 225–238 (2003)

[13] J. Estrada, et al, 'Focal plane detectors for dark energy camera.' Proc. SPIE 7735, Ground-based and Airborne Instrumentation for Astronomy III, 77351R (15 July 2010); doi: 10.1117/12.857651.

[14] The DAMIC Collaboration, "Measurement of radioactive contamination in the high-resistivity silicon CCDs of the DAMIC experiment", JINST 10 (2015) P08014.

[15] J. Estrada, J. Molina , J.J. Blostein, G. Fernandez Nuclear, "Plasma effect in silicon charge coupled devices (CCDs)", Nuclear Instruments and Methods in Physics Research A 665 (2011)

[16] J. Blostein, J. Estrada, A. Tartaglione, M. Sofo Haro, G. Fernández Moroni, Gustavo Cancelo, "Development of a novel neutron detection technique by using a boron layer coating a Charge Coupled Device", arXiv: 1408.3263 (2014).

[17] Z. Wang et al, 'A multilayer surface detector for ultracold neutrons.' NIMA 798 (2015) 30-35.

[18] W. Wei et al, "Position-sensitive detection of ultracold neutrons with an imaging camera and its implications to spectroscopy," Nuclear Instruments and Methods in Physics Research A 830 (2016) 36-43.

[19] C.L. Morris et al, "A new method for measuring the neutron lifetime using an in situ neutron detector", Review of Scientific Instruments 88, 053508 (2017).

[20] https://lansce.lanl.gov/facilities/ultracold-neutrons/index.php

[21] A. Saunders et al, "Performance of the Los Alamos National Laboratory spallation-driven solid deuterium ultra-cold neutron source," Review of Scientific Instruments 84 (2013) 013304.

[22] Karl van Bibber et al, "Quantum Sensing for High Energy Physics", arXiv:1803.11306 (2018)

[23] Stefan Baessler et al, "The GRANIT spectrometer" C. R. Physique 12 (2011) 707–728

[24] Guillaume Pignol. Probing Dark Energy Models with Neutrons. Physics [physics]. Universite Joseph Fourier, 2015. <tel-01175603>

[25] Philippe Brax, Guillaume Pignol, Damien Roulier "Probing Strongly Coupled Chameleons with Slow Neutrons" Phys. Rev. D 88, 083004 (2013)